\documentstyle[prl,aps]{revtex}
\draft
\input epsf
\input{epsf.sty}
\begin{document}

\twocolumn[\hsize\textwidth\columnwidth\hsize\csname
@twocolumnfalse\endcsname

\title{
Flux dynamics and vortex phase diagram of the new superconductor $MgB_2$
}

\author{
H. H. Wen\cite{responce}, S. L. Li, Z. W. Zhao, Y. M. Ni, Z. A. Ren, G. C. Che, and Z. X. Zhao
}

\address{
National Laboratory for Superconductivity,
Institute of Physics and Center for Condensed Matter Physics,
Chinese Academy of Sciences, P.O. Box 603, Beijing 100080, China
}

\maketitle

\begin{abstract}
Magnetic critical current density and relaxation rate have been measured on $MgB_2$ bulks from 1.6 K to $T_c$ at magnetic fields up to 8 Tesla. A vortex phase diagram is depicted based on these measurement. Two phase boundaries $H_{irr}^{bulk}(T)$ and $H_{irr}^{g}(T)$ characterizing different irreversible flux motions are found. The $H_{irr}^{bulk}(T)$ is characterized by the appearance of the linear resistivity and is attributed to quantum vortex melting induced by quantum fluctuation of vortices in the rather clean system. The second boundary $ H_{irr}^g(T) $ reflects the irreversible flux motion in some local regions due to either very strong pinning or the surface barrier on the tiny grains. 
\end{abstract}


]

The recently discovered new superconductor $MgB_2$ generates enormous interests in the field of superconductivity\cite{akimitsu}. Many important thermodynamic parameters have already been derived, such as the upper critical field $H_{c2}(0)$ = 13 - 20.4 T\cite{bud1,canfield,takano,finnemore}, the Ginzburg-Landau parameter $\kappa \approx$ 26\cite{finnemore}, and the bulk critical superconducting current density $j_c\approx 2\times 10^6 A/cm^2$ at 4.2 K and 1 T\cite{wen1} in bulk samples and $j_c\approx 8 \times 10^4 A/cm^2$ at 4.2 K and 12 T\cite{eom} in thin films. One big issue concerns however how fast the critical current will decay under a magnetic field and in which region on the field-temperature ( H-T ) phase diagram the superconductor can carry a large critical current density ( $j_c$ ). This $j_c$ is controlled by the mobility of the magnetic vortices, and vanishes at the melting point between the vortex solid and liquid. A finite linear resistivity $\rho = (E/j)_{j\rightarrow 0}$ will appear and the relaxation rate will reach 100\% at this melting point showing the starting of the reversible flux motion. In this Letter we present an extensive investigation on the flux dynamics by magnetic relaxation method. A vortex phase diagram will be depicted based on the magnetic relaxation data. 

Samples investigated here were fabricated by both high pressure ( HP ) ( P = 6 GPa at 950$^\circ$ C for 0.5 hour ) and ambient pressure ( AP ) synthesis which was described very clearly in a recent publication\cite{ren}. High pressure synthesis is a good technique for producing the $MgB_2$ superconductor since it can make the sample more dense and uniform ( in sub-micron scale ) and also prevent the oxidization of Mg element during the solid reaction. Our HP samples are very dense and look like metals with shiny surfaces after polishing. Scanning electron microscopy ( SEM ) shows that the HP sample is uniform in sub-micron scale but some disordered fine structures are found in 10 nanometer scale, being similar to the internal structure of large grains seen in the AP sample. X-ray diffraction ( XRD ) analysis on both type of samples show that they are nearly in a single phase with the second phase ( probably $MgO$ or $MgB_4$ ) less than 1 wt.\%.  For simplicity we present in this Letter only the results from the HP sample. 

Resistive transition was measured by the standard four-probe technique and the magnetic measurements were carried out by a superconducting quantum interference device ( SQUID, Quantum Design MPMS 5.5 T ) and a vibrating sample magnetometer ( VSM 8T, Oxford 3001 ). To precisely calculate the critical current density $j_c$ the sample has been cut with a diamond saw into a rectangular shape with sizes of 3.2 mm ( length ) $\times$ 2.7 mm ( width ) $\times$ 0.4 mm ( thickness ).  Fig.1 shows the diamagnetic transition of one of the samples measured in the field-cooled ( FC ) and zero-field-cooled ( ZFC ) process. In the FC process, the temperature was lowered from above $T_c$ to a desired temperature below $T_c$ under a magnetic field, and the data are collected in warming up process with field. Its signal generally describes the surface shielding current and the internal frozen magnetic flux profile. In the ZFC process, the temperature was lowered from above $T_c$ to a desired temperature below $T_c$ at a zero field and the data are collected in the warming up process with a field. Its signal generally describes the internal magnetic flux profile which is ultimately related to the flux motion. Resistive and diamagnetic measurement show that the transition is very sharp with a perfect diamagnetic signal and $T_c^{onset}$ = 40.3 K with the transition width only 0.4 K.

In Fig.1 we show the magnetization hysteresis loops ( MHL ) measured at temperatures ranging from 2 K to 38 K. The symmetric MHLs observed at temperatures up to 38 K indicate the dominance of the bulk current instead of the surface shielding current. The MHLs measured at low temperatures, such as 2 K, 4K show quite strong flux jump which will be discussed elsewhere\cite{zhaozw}. A small kink appearing in the field ascending process in low field region may be attributed to the Bean-Livingston surface barrier and will be presented separately\cite{lisl}. It is easy to see that the MHLs measured at low temperatures ( e.g., 2 K to 10 K ) are too close to be distinguishable. From these MHLs one can calculate $j_c$ via $j_c = 20 \Delta M/Va(1-a/3b)$ based on the Bean critical state model, where $\Delta M$ is the width of the MHL, V, a and b are the volume, length and width of the sample, respectively. The result of $j_c$ is
\begin{figure}[h]
    \vspace{10pt}	
    \centerline{\epsfxsize 8cm \epsfbox{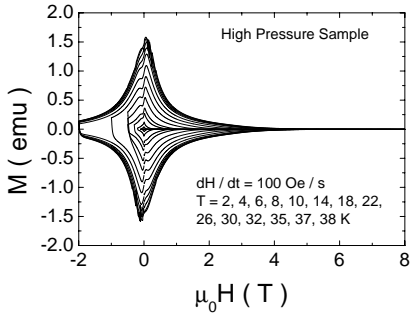}}
    \vspace{10pt}
\caption{Magnetization hysteresis loops measured at 2, 4, 6, 8, 10, 14, 18, 22, 26, 30, 32, 35, 37, K and 38 K ( from outer to inner ). All curves here show a symmetric behavior indicating the importance of bulk current instead of surface shielding current. The MHLs measured at low temperatures ( e.g., 2 K to 10 K ) are too close to be distinguishable. Strong flux jump has been observed at 2 K and 4 K near the central peak.
}
\label{fig:Fig1}
\end{figure}
\noindent  shown in Fig.2. It is clear that the bulk critical current density $j_c$ of our sample is rather high. For example, at T = 18 K and $\mu_0H$ = 1 T, we have $j_c = 1.1 \times 10^5 A / cm^2$, which is among the highest values that have been reported\cite{takano,bugoslavsky} in bulk samples. The $j_c(H)$ curves have been measured with three different field sweeping rate 200Oe/s, 100 Oe/s and 50 Oe/s. It is interesting to note that for all MHLs, there are small tails ( shown in the inset to Fig.2 ) in high field region. Accordingly the $j_c(H)$ curves show also a small tail in high field region. Since $j_c$ value in the small tail region is almost $4\times 10^4$ times smaller than that in the low field region, therefore it is safe to conclude that this small tail is due to some secondary effect, such as some local regions with very strong pinning or the surface pinning by the tiny grains. From the contour of $j_c$ vs. H shown in Fig.2 one can see that there are two ways to determine the so-called irreversibility line $H_{irr}$. The first way is to take a criterion for $j_c$ = 30 $A / cm^2$ just before the appearance of the tail. This is named as the bulk irreversibility line $H_{irr}^{bulk}(T)$ which signals the sharp drop of $j_c$ before the tail and reflects the irreversible flux motion in clean bulk samples. The second way is just to follow the small tail to a criterion of $3 A / cm^2$. In this way one can determine a higher irreversibility line $H_{irr}^{g}$ which marks a boundary above which a complete reversible flux motion occurs. As mentioned above, the small tail in our present samples is due to a secondary effect, therefore the major part of the vortex system melt at $H_{irr}^{bulk}$. Thus we may conclude that the $H_{irr}^{bulk}$ reflects the melting of the vortex matter in the clean limit, for example, in single crystals. If more pinning centers are introduced into the sample, one should be possible to push the IL more close to the upper critical 
\begin{figure}[h]
    \vspace{10pt}	
    \centerline{\epsfxsize 8cm \epsfbox{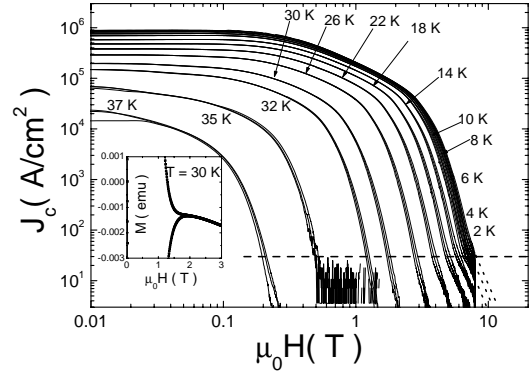}}
    \vspace{10pt}
\caption{Critical current density $j_c$ calculated based on the Bean critical state model. At each temperature the data has been measured with three field sweeping rate: 200 Oe/s, 100 Oe/s, 50 Oe/s. The faster sweeping rate corresponds to a higher dissipation and thus higher current density. From these data one can calculate the dynamical magnetic relaxation rate Q. The $j_c(H)$ curves measured at low temperatures are very close to each other showing a rather stable value of $H_{irr}^{bulk}$ when T approaches zero K. The dashed horizontal line marks the criterion of $ j_c = 30 A / cm^2 $ for determining the $H_{irr}^{bulk}$. While by following the small tail ( which has been attributed to a secondary effect ) to a criterion of $ j_c = 3 A / cm^2 $, another phase line $H_{irr}^g$ is determined.
}
\label{fig:Fig2}
\end{figure}
\noindent field $H_{c2}(T)$. The " irreversibility lines " by following these two different methods are shown in Fig.3. It is clear that the lower phase boundary $H_{irr}^{bulk}$ terminates at about 8 T at 2 K, while the extrapolated values ( $H_{irr}^{g}$ ) reaches about 12 T at 2 K. There is a large separation between these two phase boundaries. 
It is important to note that Larbalestier et al.\cite{larbalestier} have found the similar feature where they regard the lower phase line $H_{irr}^{bulk}$ as the Kramer line. Fig.3 shows also the upper critical field $H_{c2}(T)$ determined from the temperature dependent magnetization by defining the $H_{c2}(T)$ as the point at which the magnetization start to deviate from the normal state linear background\cite{wen2} and that determined from the resistive measurement by Takano et al.\cite{takano} on HP samples. A striking result is that the $H_{irr}^{bulk}(T)$ extrapolates to a rather low field, here 
for example, $H_{irr}^{bulk}(0)\approx $ 8 T, while the $H_{c2}(T)$ extrapolates to a much 
higher value ( $\approx$ 15 T )\cite{bud1} at zero K. There is a large separation between the
 two fields $H_{c2}(0)$ and $H_{irr}(0)$. If following the hypothesis of the vortex liquid above $H_{irr}$, we would conclude that there is a large region for the existence of a quantum vortex liquid at zero K. This can be attributed to a quantum fluctuation effect of vortices in bulk $MgB_2$. Although the lowest temperature in our present experiment is 1.6 K, however, from the experimental data one cannot find any tendency for $H_{irr}^{bulk}(T)$ to turn upward rapidly to meet the $H_{c2}(0)$ at zero K. One may argue that the $H_{irr}(T)$ probably can 
\begin{figure}[h]
    \vspace{10pt}	
    \centerline{\epsfxsize 8cm \epsfbox{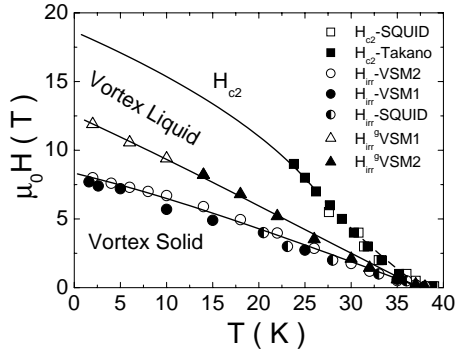}}
    \vspace{10pt}
\caption{H-T phase diagram for the new superconductor $MgB_2$. The circles represent the bulk irreversibility lines $ H_{irr}^{bulk}$ : open circles, this work measured by VSM; filled circles, another sample measured by VSM; half-filled circles, HP sample measured by SQUID. The triangles represent the $H_{irr}^g(T)$ : filled triangles, determined directly from the small tail; open triangles, following the tendency of tail ( shown by the 3 dashed lines  ) to a criterion of $ j_c = 3 A / cm^2 $. The squares represent the upper critical field $H_{c2}(T)$: filled squares, Takano's data from resistive measurement; open squares, this work from the M(T) measurement by SQUID. All solid lines are guide to the eye.}
\label{fig:Fig3}
\end{figure}
\noindent be increased to higher values by introducing more pinning centers into this sample. This is basically correct since recently a higher irreversibility line $H_{irr}(T)$ has been found in thin films\cite{eom} and bulk samples irradiated by heavy ions\cite{perkins}. This indicates that the quantum vortex melting in our present bulk sample is due to the strong quantum fluctuation which smears the perfect vortex lattice leading to the vanishing of the shear module $C_{66}$ of the vortex matter ( probably within grains ). Dense disorders will strengthen the shear module and thus enhance the bulk irreversibility line.  The irradiation of heavy ions by Perkins et al. \cite{perkins} did not suppress but strongly increase $j_c$ at a high field would suggest that the low value of $ H_{irr}^{bulk}(T) $ measured in unirradiated bulks ( like our samples here ) is not due to the weak links since otherwise the $j_c$ value would drop even faster with increasing the magnetic field after the irradiation. This indicates that the rather low $H_{irr}^{bulk}(T)$ observed in our present work reflects the melting of the vortex matter in sample with relatively pure structural details rather than the breaking of the Josephson couplings between grains. In addition, as mentioned above, the $H_{irr}^{bulk}(T)$ found in present work is very close ( or identical ) to those by other authors\cite{bugoslavsky,finnemore} found in bulks, therefore
 it shows a more intrinsic feature corresponding to the melting of vortex matter
 in the rather pure system. All these imply a strongly linked current flow between 
grains\cite{larbalestier}.

In order to investigate the flux dynamics in the vortex solid state below $H_{irr}^{bulk}$ we have carried out the dynamical 
\begin{figure}[h]
    \vspace{10pt}	
    \centerline{\epsfxsize 8cm \epsfbox{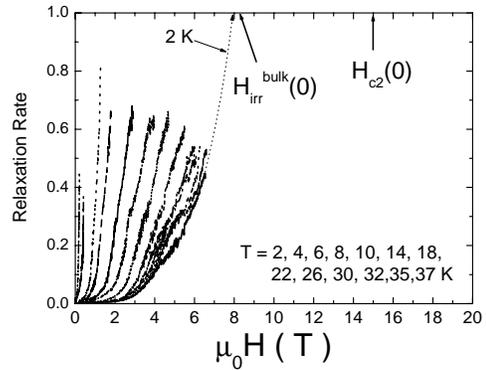}}
    \vspace{10pt}
\caption{Field dependence of the relaxation rate at temperatures of 2, 4, 6, 8, 10, 14, 18, 22, 26, 30, 32, 35, 37 K. The dashed line is a guide to the eye for 2 K. It is clear that Q will rise to 100\% at 8 T at 2 K. Since $H_{irr}^{bulk}$ is rather stable at low temperatures, it is anticipated that $H_{irr}^{bulk}(0)\approx 8 T$ being much smaller than $H_{c2}(0)\approx $ 15 T.
}
\label{fig:Fig4}
\end{figure}
\noindent relaxation measurement. According to Schnack et al.\cite{schnack} and Jirsa et al.\cite{jirsa}, in a field sweeping process, if the field sweeping rate is high enough, the quantity $Q = dln\Delta M/dln(dH/dt)$ is identical to the relaxation rate $S =- dlnM/dlnt$ measured in the conventional relaxation method, where Q is called as the dynamical relaxation rate, $\Delta M$ the width of the MHL, $dH/dt$ the field sweeping rate. The raw data with three different sweeping rate ( 200 Oe/s, 100 Oe/s, and 50 Oe / s ) are shown in Fig.2. The Q values vs. field for different temperatures are determined and shown in Fig.4. It is clear that the relaxation rate increases monotonically with the external magnetic field and extrapolates to 100\% at the bulk melting point $H_{irr}^{bulk}$. A closer inspection to the data finds that the rising of the relaxation rate Q vs. field is slowered down when it runs into the small tail region. Since the small tail is a secondary effect as mentioned before, we leave it to a future investigation. Here we concentrate on the flux dynamics before the setting in of the small tail. At 2 K it is found from the Q(H) data that the melting field is about 8 T, being very close to $H_{irr}^{bulk}( T = 2 K )$ determined from the $j_c(H)$ curve. It is known that the $H_{irr}^{bulk}(T)$ is rather stable in low temperature region, therefore we can anticipate a rather low value of 
$H_{irr}^{bulk}(0)$ which is below 9 T being much lower than $H_{c2}(0)$. The large separation between $H_{irr}^{bulk}(0)$ and $H_{c2}(0)$ may manifest the existence of the quantum vortex liquid due to strong quantum fluctuation of vortices in the pure system of $MgB_2$. Fig.5 shows the temperature dependence of the relaxation rate Q. The arrows point at the irreversibility temperatures at the corresponding fields $H_{irr}^{bulk}(T)$. It is clear that the relaxation rate in wide temperature region keeps rather stable against the thermal activation and fluctuation. However, when the bulk melting point $H_{irr}^{bulk}(T)$ is approached the relaxation 
\begin{figure}[h]
    \vspace{10pt}	
    \centerline{\epsfxsize 8cm \epsfbox{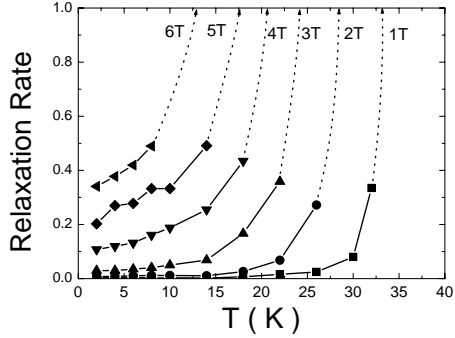}}
    \vspace{10pt}
\caption{Temperature dependence of the relaxation rate at fields of 1, 2, 3, 4, 5, 6 T. The dashed lines are guides to the eye. The arrows point at the bulk irreversibility temperatures $T_{irr}^{bulk}$ determined from the $ j_c(H) $ curve.
}
\label{fig:Fig5}
\end{figure}
\noindent rate will quickly jump to 100\%. This may indicate that the thermal fluctuation is not the dominant  process for the flux depinning in the new superconductor $MgB_2$. It shows high possibility for the vortex quantum melting even at a finite temperature. Worthy to note is that the quantum tunneling rate is extremely low at an intermediate field, such as Q = 0.7\% at 2 K and 2 T, but is rather high at a high field, for example Q = 33\% at 2 K and 6 T. This may imply that the field will greatly enhance the vortex quantum fluctuation and tunneling.

In conclusion, in rather pure samples of $MgB_2$, the irreversibility field is rather low comparing to the upper critical field in low temperature region showing the possible existence of the quantum vortex liquid due to strong quantum fluctuation. The temperature and field dependence of the relaxation rate may further suggest that the vortex melting at a finite temperature is also induced by strong quantum fluctuation in pure systems, such as single crystals and bulks. A small tail following the $j_c(H)$ curve in high field region is attributed to a secondary effect induced by strong pinning or surface barrier on some tiny grains. This opens a way for the higher irreversibility line when more pinning centers are introduced into the pure sample. 

\acknowledgements
This work is supported by the National Science Foundation of China (NSFC 19825111) and the Ministry of Science and Technology of China ( project: NKBRSF-G1999064602 ). HHW gratefully acknowledges Prof. B. Ivlev and Dr. A. F. Th. Hoekstra for fruitful discussions, and continuing financial support from the Alexander von Humboldt foundation, Germany.

\end{document}